# Thickness dependence of the anomalous Nernst effect and the Mott relation of Weyl-semimetal Co$_2$MnGa thin films

*Gyu-Hyeon Park,*[1,2] *Helena Reichlova,*[3] *Richard Schlitz,*[3,4] *Michaela Lammel,*[1,4] *Anastasios Markou,*[5] *Peter Swekis,*[5] *Philipp Ritzinger,*[3] *Dominik Kriegner,*[5,6] *Jonathan Noky,*[5] *Jacob Gayles,*[5] *Yan Sun,*[5] *Claudia Felser,*[5] *Kornelius Nielsch,*[1,2] *Sebastian T. B. Goennenwein,*[3,4] *Andy Thomas*[1,4,*]

[1]*Leibniz Institute for Solid State and Materials Research Dresden (IFW Dresden), Institute for Metallic Materials, Helmholtzstrasse 20, 01069 Dresden, Germany*
[2]*Technische Universität Dresden, Institute of Materials Science, 01062 Dresden, Germany*
[3]*Institut für Festkörper- und Materialphysik and Würzburg-Dresden Cluster of Excellence ct.qmat, Technische Universität Dresden, 01062 Dresden, Germany*
[4]*Center for Transport and Devices of Emergent Materials, Technische Universität Dresden, 01062 Dresden, Germany*
[5]*Max Planck Institute for Chemical Physics of Solids, Nöthnizer Str. 40, 01187 Dresden, Germany*
[6]*Institute of Physics ASCR, v.v.i., Cukrovarnicka 10, 162 53, Praha 6, Czech Republic*

**Abstract**

We report a robust anomalous Nernst effect in Co$_2$MnGa thin films in the thickness regime between 20 and 50 nm. The anomalous Nernst coefficient varied in the range of -2.0 to -3.0 µV/K at 300 K. We demonstrate that the anomalous Hall and Nernst coefficients exhibit similar behavior and fulfill the Mott relation. We simultaneously measure all four transport coefficients of the longitudinal resistivity, transversal resistivity, Seebeck coefficient, and anomalous Nernst coefficient. We connect the values of the measured and calculated Nernst conductivity by using the remaining three magneto-thermal transport coefficients, where the Mott relation is still valid. The intrinsic Berry curvature dominates the transport due to the relation between the longitudinal and transversal transport. Therefore, we conclude that the Mott relationship is applicable to describe the magneto-thermoelectric transport in Weyl semimetal Co$_2$MnGa as a function of film thickness.

---

[*] a.thomas@ifw-dresden.de

## I. INTRODUCTION

The anomalous Nernst effect (ANE) experimentally manifests as an anomalous transversal voltage, that is perpendicular to both the heat current and magnetization. It represents the thermoelectric counterpart of the anomalous Hall effect (AHE). The origin of both the AHE and ANE have contributions from the extrinsic and intrinsic (i.e. Berry phase) mechanisms [1- 3]. When the AHE and the ANE effects are dominated by the momentum space Berry curvature of the electronic structure, it allows for an advantageous prediction of compounds with large electrical and thermal effects based solely on the topology of the electronic structure [4-6].

The AHE and the ANE are connected via the Mott relation [2, 3]. In general, the Seebeck coefficient can be expressed via the Mott relation: $S = \frac{\pi^2 k_B^2 T}{3e\sigma}(\frac{\partial \sigma}{\partial E})_{E_F}$, where $k_B$ is the Boltzmann constant, $e$ is the elementary charge, and the energy (E) derivative of the electrical conductivity ($\sigma$) at the Fermi level ($E_F$) [6, 7]. Generally, the Mott relation can be applied in materials where each charge carrier acts independently [8, 9]. Additionally, the Mott relation holds to explain the dominant intrinsic character in the anomalous transport in ferromagnetic materials such as the spinel feature of CuCr$_2$Se$_{4-x}$Br$_x$ [10], diluted magnetic semiconductors (DMS) [11], and with Berry phase or curvature [6, 12, 13]. However, recent theoretical work [12] questions the validity of the Mott relation in materials with a nontrivial topology of the electronic bands and suggests that the ANE could be sensitive to electronic states invisible to the AHE. The relationship between the ANE and the AHE was systematically studied and confirmed by Pu *et al.* [11] in the ferromagnetic semiconductor GaMnAs, in which the AHE arise from the intrinsic spin-orbit coupling. A systematic study of the Mott relation in thin films with a nontrivial topology of the electronic states is missing, mostly because, typically the ANE coefficient is small and not easy to measure and quantify. In addition, a series of samples is required since a simple temperature dependent measurement of the nontrivial AHE and ANE in one sample is not conclusive.

$Co_2MnGa$, which is a member of the $Co_2YZ$-based full Heusler family, has very interesting properties, such as a high Curie temperature of 700 K and a high spin polarization [14]. Furthermore, $Co_2MnGa$ has been regarded as a promising material, since it is a magnetic Weyl semimetal and has an unconventional topological surface state [15-17]. Additionally, recent work has identified the Berry curvature as the origin of the large AHE and suggested that the topology can be tuned by selecting the magnetic space group [1]. More recently, Belopolski *et al.* systematically measured the anomalous Hall response by considering the Berry curvature field and linked it with the evaluated topological Weyl fermion lines [18]. Experimentally, the magneto-thermoelectric properties of bulk $Co_2MnGa$ have demonstrated a record value for the anomalous Nernst coefficient (ANC or $S_{xy}$) of -6 μV/K [6]. Additionally, we reported an $S_{xy}$ of approximately -2 μV/K at 300 K [19] in a 50 nm thin film of $Co_2MnGa$.

Here, we report a systematic study of the ANE in $Co_2MnGa$ thin films that exhibit large $S_{xy}$ values. We confirm the robustness of the reported value above -2 μV/K in the thickness range of 20 to 50 nm. We further employ the thickness series to systematically study the relationship between the ANE and AHE. We reveal the validity of the Mott relation in this particular material at 300 K by comparing the measured values of the anomalous Nernst conductivity with the calculated values.

## II. EXPERIMENT, RESULTS AND DISCUSSION

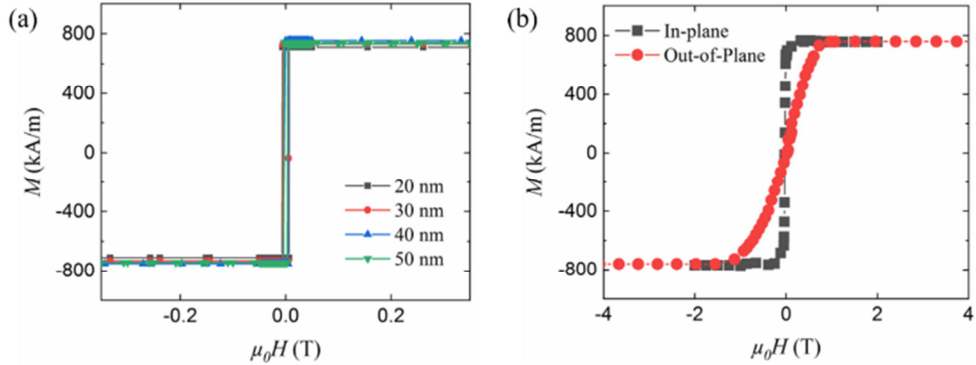

FIG. 1 (a) In-pane magnetization of $Co_2MnGa$ thin films with thicknesses between 20 nm and 50 nm. All thicknesses exhibit saturated magnetization of approximately 720 kA/m. (b) comparison of in-plane and out-of-plane magnetization hysteresis loops of the 40 nm $Co_2MnGa$ film measured at 300 K.

$Co_2MnGa$ thin films were deposited with different thicknesses (20, 30, 40 and 50 nm), by magnetron sputtering on MgO (001) substrates using a multisource Bestec UHV deposition system. We deposited 3 nm of Al at room temperature to prevent oxidation. The thin films were post annealed at 500°C. The chemical composition and structural investigation conducted by X-ray techniques revealed $Co_2MnGa$ Bragg peaks [20] revealing high degree of atomic order, consistent with previous work [19]. Unpatterned material was studied by SQUID magnetometry: All thicknesses exhibit saturated magnetization values of approximately 720 kA/m and similar coercive fields. Furthermore, we show typical in-plane and out-of-plane magnetization hysteresis loops for 40 nm $Co_2MnGa$ measured at 300 K depicted in Fig. 1(a) and (b), respectively.

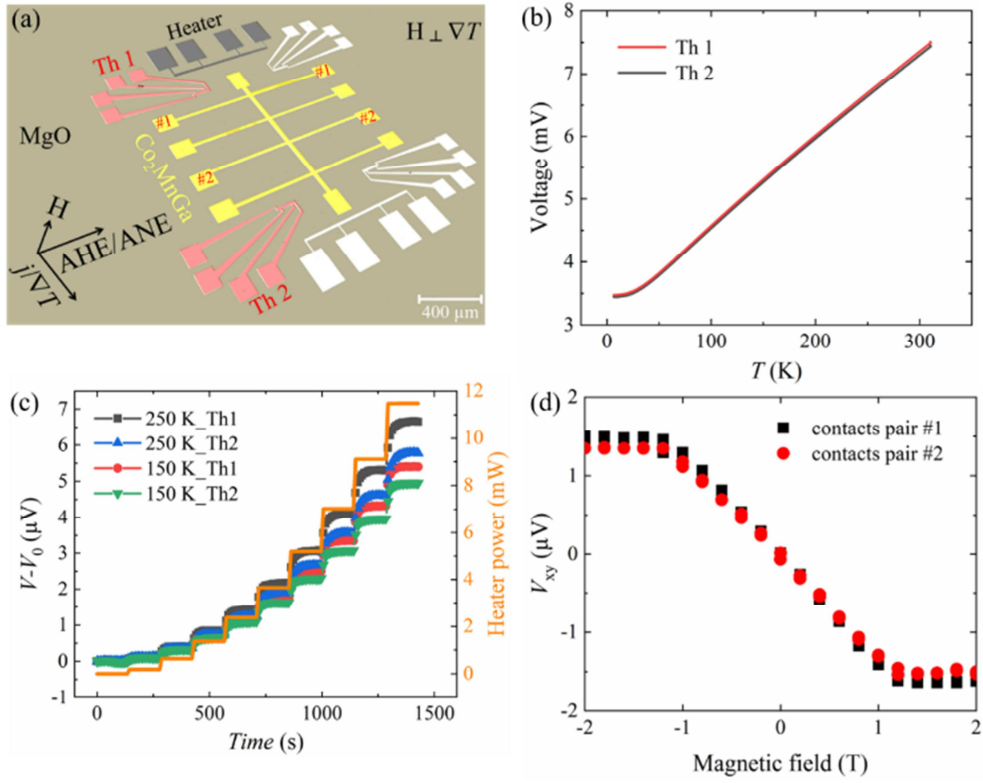

FIG. 2 Optical microscopy and thermal gradient evaluation. (a) False color optical microscopy image of our measurement platform. (b) The calibration curves of Th1 and Th2 measured at the corresponding pairs during homogeneous heating of the sample. (c) Voltages measured at the two thermometers at different heater power at two specific base temperatures of 150 K and 250 K. (d) Homogeneity of the thermal gradient measured at two pairs of contacts.

The films were patterned into Hall bars by optical lithography and by a combination of HCl and Ar/$O_2$ plasma etching. The Fig. 2(a) shows the full devices structure. After etching, the heater and thermometers were defined in a lift-off process with 30 nm of sputtered Pt. The temperature gradient was generated by an on-chip heater, and the temperature on the cold and hot sides of the sample was monitored by two on-chip thermometers (Th1 and Th2). An external magnetic field was applied perpendicular to the sample plane. An Oxford Instruments cryostat with two thermometers is used to monitor the sample's base temperature. We define the thermal gradient as $\nabla T = \Delta T/L$, where $\Delta T$ is the temperature difference and $L$ is the distance between the on chip thermometers Th1 and Th2 (1.3 mm). The measured voltage at each thermometer as a function of the base temperature is shown in Fig. 2(b). This serves as a calibration curve for the evaluation of the thermal gradient. Fig. 2(c) depicts the measured voltage from Th1 and Th2 at temperatures of 250 K and 150 K while the heater power is increased stepwise over time. Note that we measured the same sequence at all temperatures. Then, we evaluated $\nabla T$ to determine $S_{xy}$, since thermal conductivity of the $MgO_x$ substrate can vary with the base temperature. As a consequence, we determined the typical $\nabla T$ and the error of the thermal gradient evaluation. Notably, the two anomalous Nernst voltage values ($V_{xy}$) measured at the two pairs of contacts #1 and #2 (Fig. 2(a)) show only 7% difference indicating a homogenous thermal gradient in Fig. 2(d). The evaluation process is similar to our previous work, where we demonstrated that the thermal gradient along the y-direction is negligible [19, 21].

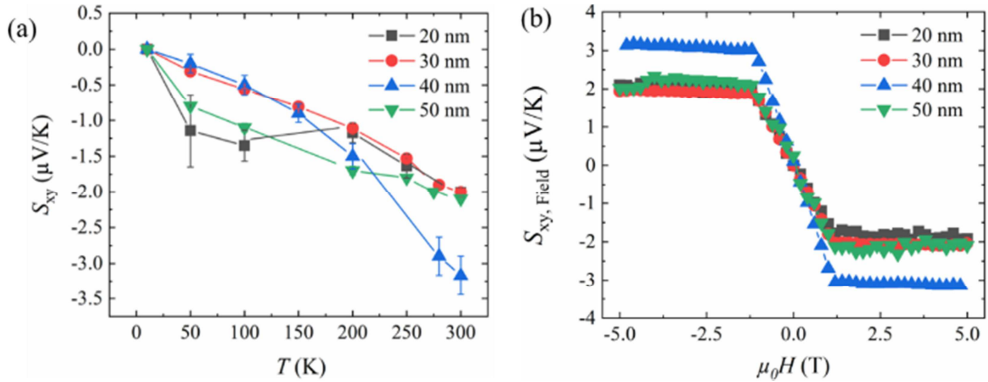

FIG. 3 Anomalous Nernst measurement. (a) Temperature dependence of $S_{xy}$ evaluated for each sample thickness. (b) Magnetic field dependence of the $S_{xy,Field}$ at 300 K.

To quantify $S_{xy}$, we swept the external magnetic field from –5 to 5 Tesla and measured the $S_{xy}$ at the transverse contacts #1, Please note that the difference of the measured value between contact #1 and #2 is negligible (~0.13 µV/K). The ANC of all thicknesses of the Co$_2$MnGa thin films at various temperatures are presented in Fig. 3(a). The ordinary Nernst effect is very small in Co$_2$MnGa thin films. Consequently, $S_{xy,\text{Field}}$ at 300 K shows only a small change after saturation (Fig. 3(b)). Therefore, the coefficient $S_{xy}$ was evaluated in the following way: $E_{\text{ANE}} = -S_{xy} m \times \nabla T$, where $E_{\text{ANE}}$ and $m$ are the electric field induced by the ANE and the magnetization vector, respectively. Before applying $\nabla T$, the base temperature of the setup was obtained at a specific temperature. Then, a current was applied to a heater to generate a temperature gradient. Two thermometers were employed to quantify $\nabla T$, afterwards, the $S_{xy}$ values are evaluated. In the measured temperature range from 10 K to 300 K, $S_{xy}$ gradually increases with increasing temperatures, as expected for a magneto-thermal effect far below the Curie temperature [22]. All samples in the thickness range between 20 to 50 nm exhibit large $S_{xy}$ above -2 µV/K. Interestingly, the 40 nm films exhibit even higher values of -3 µV/K.

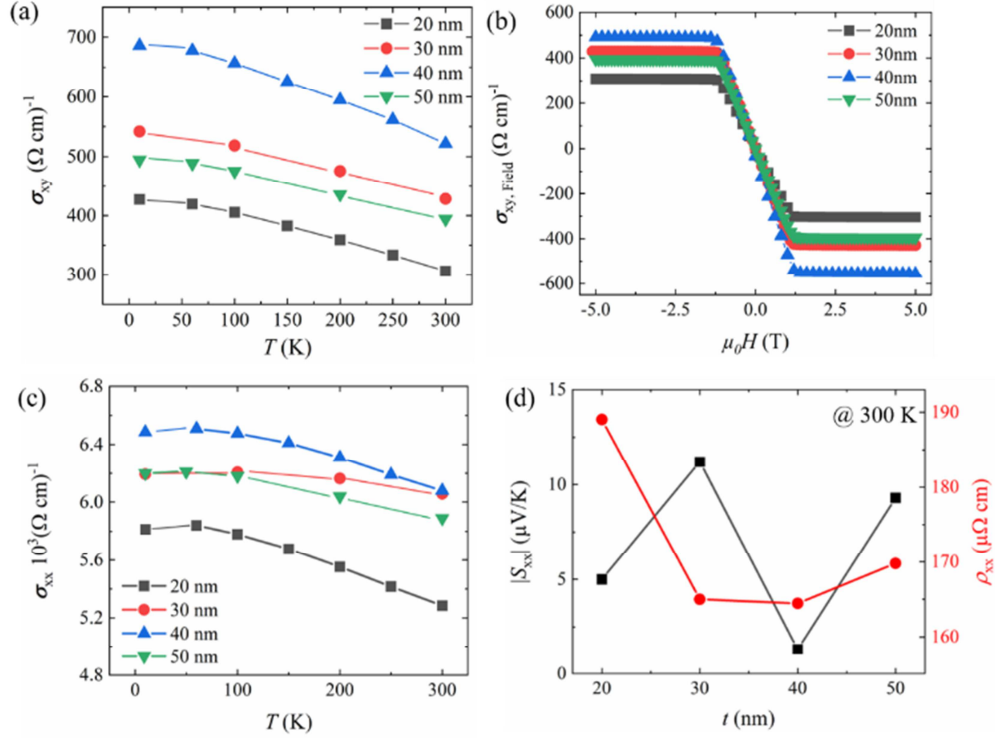

FIG. 4 (a) Temperature dependence of $\sigma_{xy}$ evaluated for each sample thickness. (b) Magnetic field dependence of the $\sigma_{xy,Field}$ at 300 K. (c) Longitudinal conductivity $\sigma_{xx}$ of each thickness as a function of temperature. (d) Thickness dependence of Seebeck coefficient $S_{xx}$ and Longitudinal resistivity $\rho_{xx}$ measured at 300 K.

We evaluated the longitudinal conductivity as $\sigma_{xx} = 1/\rho_{xx}$ and the anomalous Hall conductivity (AHC) as $\sigma_{xy} = -\rho_{xy}/(\rho_{xx}^2 + \rho_{xy}^2)$, where $\rho_{xx}$ is the longitudinal and $\rho_{xy}$ is the transversal resistivity [2, 22, 23]. The small decrease in $\sigma_{xy}$ with increasing temperature is caused by the Hall resistivity slowly changing with linear Hall coefficient as $R_0$ [1]. Similarly, $\sigma_{xy}$ gradually decreased with increasing temperature from 10 K to 300 K, as shown in Fig. 4(a). Fig. 4(b) shows $S_{xy}$ as a function of the applied external field $\mu_0 H$. The saturation field at 300 K is 1.4 T. The ANE and AHC are both weakly reduced in the 50 nm $Co_2MnGa$ films, which could be related to the observations reported in Ref. 17 and 24. As an example, Chuang *et al.* suggested that the intrinsic mechanism dominates at lower thicknesses. In this study, the ANE of Ni thin films increased to 25 nm and decreased at higher thicknesses, revealing that the enhancement of the ANE in thin films is mostly dominated by the intrinsic and side-jump mechanisms [25]. Notably, thickness variations of the ANE cannot be explained by the variation in the magnetization, as shown in Fig. 1(a). As already presented, the saturation magnetization of all studied samples is very similar. The longitudinal conductivity $\sigma_{xx}$ is shown in Fig.4 (c). In Fig. 4(d), we quantified the Seebeck coefficient ($S_{xx}$) using $S_{xx} = V_{xx}/\nabla T$, where $V_{xx}$ is the Seebeck voltage at 300 K and measured the longitudinal resistivity $\rho_{xx}$. The variation of $\sigma_{xx}$ and $S_{xx}$ between various samples can be caused by a weak stoichiometry variation, as shown for example by Sato *et al.* [26]. In Table I, we summarized the $|\sigma_{xx}|$, $|\sigma_{xy}|$, $S_{xx}$, $|\rho_{xx}|$, and $|\rho_{xy}|$ with thickness from 20 to 50 nm.

Table I. Summary of resistivity and conductivity of longitudinal and transversal, and Seebeck coefficient measured at 300 K.

| Thickness [nm] | $|\sigma_{xx}|$ $10^3[\Omega\ cm]^{-1}$ | $|\sigma_{xy}|$ $10^3[\Omega\ cm]^{-1}$ | $S_{xx}$ [μV/K] | $|\rho_{xx}|$ [μΩ cm] | $|\rho_{xy}|$ [μΩ cm] |
|---|---|---|---|---|---|
| 20 | 5.290 | 0.306 | 5.0 | 189.0 | 11.0 |
| 30 | 6.060 | 0.425 | 11.2 | 165.0 | 12.0 |
| 40 | 6.080 | 0.518 | 1.3 | 164.5 | 14.1 |
| 50 | 5.890 | 0.392 | 9.3 | 169.8 | 11.3 |

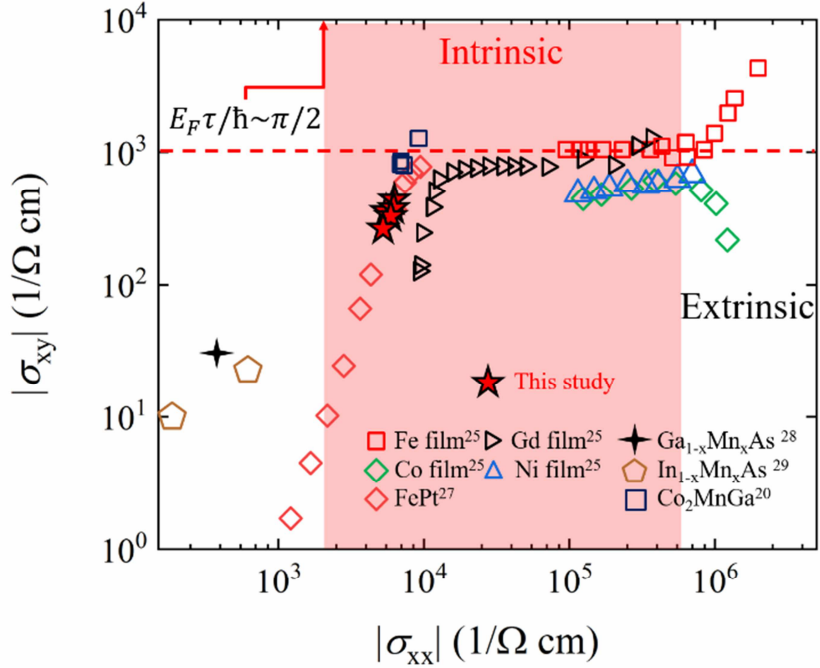

FIG. 5 Summary of the absolute values of $\sigma_{xy}$ as a function of $\sigma_{xx}$ (determined experimentally) in various thin film materials. The data are taken from T. C. Chuang *et al.* (Ref. 25) for Fe, Gd, Co, and Ni thin film; from A. Markou *et al.* (Ref. 20) for $Co_2MnGa$; from Y. M. Lu *et al.* (Ref. 27) for FePt; from K. W. Edmonds *et al.* (Ref. 28) for $Ga_{1-x}Mn_xAs$; from A. Oiwa *et al.* (Ref. 29) for $In_{1-x}Mn_xAs$.

In the following we discuss the character of the AHE. Different contributions to the AHE can be studied by estimating the dependency between the longitudinal and transversal conductivity [23]. This approach was employed to study many materials in order to distinguish intrinsic and extrinsic contributions to the AHE [7]. In the simplest scenario, three regions can be identified as illustrated in Fig. 5. First, in the poorly conducting regime ($\sigma_{xx} < \sim 3 \times 10^3 \, \Omega^{-1} cm^{-1}$), the dependence of $\sigma_{xy}$ on the residual resistivity is well described by $\sigma_{xy}^{AHE} \propto \sigma_{xx}^{1.6}$ (experimentally) [7, 30]. Secondly, in the intermediate region ($\sigma_{xx} \sim 3 \times 10^3 - 5 \times 10^5 \Omega^{-1} cm^{-1}$), the behavior can be explained by the intrinsic Berry-phase contribution [31]. For Co$_2$MnGa thin films, the combination of $\sigma_{xx}$ and $\sigma_{xy}$ is located in this region. Accordingly, the AHE in Co$_2$MnGa thin films may include a Berry curvature contribution. Thirdly, in the extremely conducting case ($\sigma_{xx} \geq 5 \times 10^6 \Omega^{-1} cm^{-1}$), $\sigma_{xy}$ depends on the constituents of the compounds and on Landau-level formation at low magnetic field due to the high mobility of the charge carriers [31].

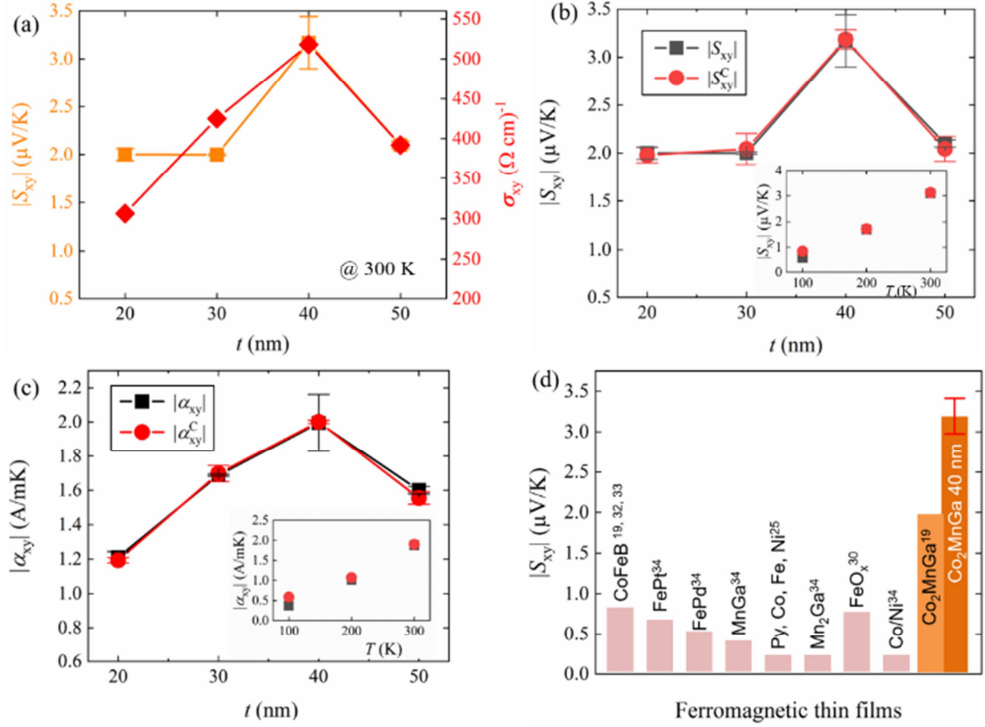

FIG. 6 (a) Thickness dependence of the anomalous Nernst effect and anomalous Hall effect measured at 300 K in Fig. 6 (a). Measured and calculated thickness dependence of Nernst coefficient ($|S_{xy}|$ and $|S_{xy}^C|$) (b), and conductivity ($|\alpha_{xy}|$ and $|\alpha_{xy}^C|$) (c) for all samples at 300 K. The insets show the temperature dependence of $|S_{xy}|$ and $|\alpha_{xy}|$ in (b) and (c), respectively. Both insets depict the values for the 40 nm sample of Co$_2$MnGa. The calculated red circled symbols are the best fits using Eq. (1) and (2) with $n = 1.82 \pm 0.08$ for $S_{xy}^C$ and $n = 1.86 \pm 0.05$ for $\alpha_{xy}^C$ respectively. (d) Comparison of the $|S_{xy}|$ values of various ferromagnetic thin films and the current study at 300K. The data are taken from H. Reichlova, J. Wells, and S. Tu *et al.* (Ref. 19, 32, 33) for $Co_2MnGa$ thin film; from T. C. Chuang *et al.* (Ref. 25) for Py, Co, Fe, and Ni thin film; from R. Ramos *et al.* (Ref. 30) for $FeO_x$ thin film, from K. Hasegawa *et al.* (Ref. 34) for FePt, FePd, MnGa, $Mn_2Ga$, and Co/Ni thin film.

Our series represents a good model system to study the Mott relation, because it has a large ANE. In addition, the ANE exhibits an anomaly as a function of the thickness although the magnetization is identical for all studied samples. Specifically, most of the samples have similar ANE values of ~-2 µV/K (20, 30, and 50 nm), but one value is larger than the others, i.e. ~ -3 µV/K (40 nm). These features are repeated in both the ANE and the AHC measured at 300 K, as shown in Fig. 6(a). Previous studies correlated the large ANE in $Co_2MnGa$ with the intrinsic Berry curvature. Our study demonstrates that the ANE and the AHE exhibit comparable trends in the full thickness range as can be seen in Fig.6 (a) by direct comparison between the ANE and the AHE. A more accurate test of the Mott relation, however, requires knowledge of all four magneto-thermal transport coefficients and it will be presented in the following paragraphs. In the end, we provide a summary of the $|S_{xy}|$ values of various ferromagnetic thin films with thickness below 1 µm in Table II.

Table II. Overview of $|S_{xy}|$ values for various ferromagnetic thin films.

| Material | $|S_{xy}|$ [µV/K] | Thickness [nm] | Ref |
|---|---|---|---|
| CoFeB | 0.8 | 15 | 19, 32, 33 |
| FePt | 0.65 | 30 | 34 |
| FePd | 0.5 | 30 | 34 |
| Co/Ni | 0.2 | 30 | 34 |
| $Mn_2Ga$ | 0.2 | 30 | 34 |
| $L1_0$-MnGa | 0.2 | 30 | 34 |
| $FeO_x$ | 0.74 | 50 | 30 |
| Py, Co, Fe, Ni | 0.06 | 1-100 | 25 |
| $Co_2MnGa$ | 2 | 20, 50 | 19 |
| $Co_2MnGa$ | 3.17 ± 0.27 | 40 | This work |

We employed an approach described by Guin and Pu *et al.* [6, 11]; because the $S_{xy}$ is related to other transport coefficients, we can determine the transverse thermoelectric conductivity as $\alpha_{yx}(=-\alpha_{xy})$ by

$$\alpha_{yx} = \sigma_{yx}S_{xx} + \sigma_{xx}S_{yx} \tag{1}$$

where $S_{yx}(=-S_{xy})$. On the other hand, $\alpha_{xy}$ and $S_{xy}$ can be calculated under the assumption that the Mott relation is valid. The measured values of perpendicular ($\rho_{xx}$) as well as transversal resistivity ($\rho_{xy}$), and $S_{xx}$ are employed to estimate $|S_{xy}^c|$ and $|\alpha_{xy}^c|$ as:

$$|S_{xy}^c| = \left|\frac{\rho_{xy}}{\rho_{xx}}\left(T\frac{\pi^2 k_B^2}{3e}\frac{\lambda'}{\lambda} - (n-1)S_{xx}\right)\right| \tag{2},$$

and

$$|\alpha_{xy}^c| = \left|\frac{\rho_{xy}}{\rho_{xx}^2}\left(T\frac{\pi^2 k_B^2}{3e}\frac{\lambda'}{\lambda} - (n-2)S_{xx}\right)\right| \tag{3}$$

We plotted these calculations together with the experimental values, such as $|S_{xy}|$ and $|\alpha_{xy}|$ in Fig. 6 (b) and (c). By fitting Eq. (2) to $|S_{xy}|$ [Fig. 6(b)] or Eq. (3) to $|\alpha_{xy}|$ [Fig. 6(c)] for all samples and temperatures, we can determine each pair of values for *n* and $\lambda'/\lambda$ ($n = 1.82 \pm 0.08$, $\lambda'/\lambda = 3.25 \pm 0.05 \times 10^{19}$ $J^{-1}$ for $S_{xy}^c$, $n = 1.86 \pm 0.05$, $\lambda'/\lambda = 3.25 \pm 0.00 \times 10^{19}$ $J^{-1}$ for $\alpha_{xy}^c$) for all the presented data. These *n* values correspond to a dominantly intrinsic character of the AHE and leads to an excellent agreement between measured and calculated $\alpha_{xy}^c$ for the majority of thicknesses. The presented analysis demonstrates the applicability of the Mott rule to the studied systems and the common physical origin behind the ANE and the AHE in the Weyl semimetal Co$_2$MnGa. In the end, we provide a summary of the $|S_{xy}|$, $|S_{xy}^C|$, $|a_{xy}|$, and $|a_{xy}^C|$ values of Co$_2$MnGa thin films with thickness series in Table III.

Table III. Summary of measured and calculated Anomalous Nernst value and transverse thermoelectric conductivity at 300 K.

| Thickness [nm] | $|S_{xy}|$ [μV/K] | $|S_{xy}^C|$ [μV/K] | $|a_{xy}|$ [A/mK] | $|a_{xy}^C|$ [A/mK] |
|---|---|---|---|---|
| 20 | 2.00 ± 0.06 | 1.98 ± 0.08 | 1.21 ± 0.03 | 1.19 ± 0.02 |
| 30 | 2.00 ± 0.01 | 2.04 ± 0.16 | 1.69 ± 0.01 | 1.70 ± 0.05 |
| 40 | 3.17 ± 0.27 | 3.19 ± 0.10 | 2.00 ± 0.16 | 2.00 ± 0.01 |
| 50 | 2.10 ± 0.04 | 2.04 ± 0.13 | 1.60 ± 0.02 | 1.55 ± 0.04 |

To date, many studies have been conducted to search for the origin of the AHE and the ANE [1, 5]. High $S_{xy}$ values were achieved by tuning the chemical composition [35, 36] and band structure [6] and predicted in simulations [37]. These studies were not only motivated by the understanding of the nature of the unconventional topological states and their consequences for magneto-thermal transport but were also motivated by a search for a path towards spin-caloritronic devices [32, 38]. Recently, the origin of this anomaly was explained by topological Weyl fermion lines in the Berry curvature in bulk systems [18]. However, extrinsic scattering effects are expected to be small in the thin film regime, because of the contributions from the surface states. Therefore, we believe it is justified to take the intrinsic contributions into account. Notably, the present study offers a promising thin film material with a large room temperature $S_{xy}$, that is robust over a thickness range of 20 to 50 nm. We present various $S_{xy}$ values for ferromagnetic thin films in Fig. 6(d). Therefore, it can be seen that our Co$_2$MnGa thin films have outstanding $S_{xy}$ values and represent a record value within the ferromagnetic thin film experiments.

### III. SUMMARY

In summary, we studied the anomalous Nernst effect in a thickness series of $Co_2MnGa$ and confirmed the record large value above -2 µV/K in the thickness range of 20 to 50 nm. Combining the electrical and the thermoelectric measurements and the extracted Nernst coefficients, we employed this system to study the Mott relation in thin films with a nontrivial topology inherent to band structure. We observed that the ANE is largest -3 µV/K for the 40 nm thin film. An analog trend is observed when studying the AHE. By comparing various thicknesses with magnetometry measurements, we observe that this trend is independent in the variation of magnetization. In agreement with recent reports, we believe that both the ANE and the AHC have contributions arising from not only finite Berry phase curvature. Furthermore in this thinner regime, intrinsic behavior plays a dominant role. We show that the Mott relation is valid in this material with a nontrivial topology of the band structure.

**ACKWNOWLEDGMENTS**

We thank the EU via the FET Open RIA Grant no. 766566. The DFG through SFB 1143 (project-id 247310070) and the Würzburg-Dresden Cluster of Excellence on Complexity and Topology in Quantum Matter – ct.qmat (EXC 2147, project-id 39085490). Furthermore, we thank the DFG via the CRC 1143/C08. D.K. acknowledges the support by the Operational Program Research, Development and Education financed by European Structural and Investment Funds and the Czech Ministry of Education, Youth and Sports (Project No. CZ.02.2.69/0.0/0.0/16_027/0008215), and Ministry of Education of the Czech Republic Grant No. LM2018110 and LNSM-LNSpin, Czech Science Foundation Grant No. 19-28375X